\documentclass[aps,prl,preprint,tightenlines,superscriptaddress,showpacs]{revtex4}
\def\mystyle{4}

\if\mystyle4
 \def\figonescale{0.95}
 \def\figtwoscale{0.95}
 \def\figone{fig1_mbcfitcol.eps}
 \def\figtwo{fig2_q2col.eps}
  \def\bellelogo{\vbox to 16mm{
                 \vss\hbox{\resizebox{!}{3cm}{
                 \includegraphics{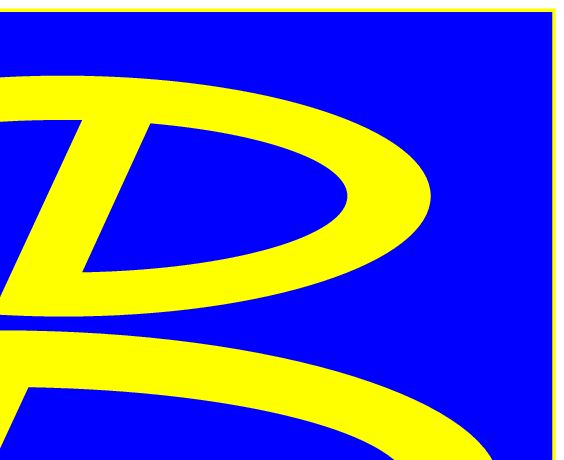}}}}\vspace{-1cm}}
  \def\preprintA{\hbox{\hfil Belle Preprint 2003-13}}
  \def\preprintB{\hbox{\hfil KEK Preprint 2003-45}}
  \def\preprintC{\hbox{\hfil DPNU-03-18}}
  \def\SomeSpaceIfPreprint{\quad\\[1cm] \Large}
\else
 \def\figonescale{0.45}
 \def\figtwoscale{0.5}
 \def\figone{fig1_mbcfitmono.eps}
 \def\figtwo{fig2_q2mono.eps}
 \def\bellelogo{}
 \def\SomeSpaceIfPreprint{}
 \def\preprintA{}
 \def\preprintB{}
 \def\preprintC{}
\fi

\if\mystyle2
  \def\mydate{\date{September 12, 2003, Version 0912, option 4}}
\else
  \def\mydate{\date{September 12, 2003}}
\fi

\usepackage{graphicx}
\def\ResultBrBtoKll{(4.8^{+1.0}_{-0.9} \pm 0.3 \pm 0.1) \times 10^{-7}}
\def\ResultBrBtoKstarll{(11.5^{+2.6}_{-2.4} \pm 0.8 \pm 0.2 ) \times 10^{-7}}

\def\epem{e^+e^-}
\def\mumu{\mu^+\mu^-}
\def\elel{\ell^+\ell^-}

\def\qqbar{q\bar{q}}

\def\KP{K^+}
\def\piP{\pi^+}
\def\piM{\pi^-}
\def\piZ{\pi^0}
\def\KZ{K^0}
\def\KS{K^0_S}

\def\Kst{K^*}
\def\KstZ{K^{*0}}
\def\KstP{K^{*+}}

\def\Kll{K\elel}
\def\Kstarll{\Kst\elel}

\def\BtoXsgamma{B\to X_s\gamma}
\def\BtoKstargamma{B\to K^*\gamma}
\def\BtoKorKstarPZ{B\to K^{(*)}\piZ}
\def\Xs{X_s}
\def\BtoKll{B\to K\ell^+\ell^-}

\def\BtoKstarll{B\to K^*\ell^+\ell^-}
\def\BtoKstaree{B\to K^*\epem}
\def\BtoKstarmumu{B\to K^*\mumu}
\def\BtoKorKstarll{B\to K^{(*)}\ell^+\ell^-}
\def\BtoKorKstaremu{B\to K^{(*)}e^\pm\mu^\mp}
\def\BtoXsll{B\to\Xs\ell^+\ell^-}

\def\BtoKSll{B^0\to\KS\elel}
\def\BtoKPll{B^+\to\KP\elel}
\def\BtoKstarZll{B^0\to K^{*0}\elel}
\def\BtoKstarPll{B^+\to K^{*+}\elel}

\def\JpsiKorKstar{J/\psi\,K^{(*)}}
\def\BtoJpsiK{B\to J/\psi\,K}

\def\GeV{{\rm~GeV}}
\def\GeVc{{\rm~GeV}/c}
\def\GeVcc{{\rm~GeV}/c^2}
\def\MeV{{\rm~MeV}}

\def\MeVcc{{\rm~MeV}/c^2}

\def\fbinv{{\rm~fb}^{-1}}
\def\mrad{{\rm~mrad}}

\def\Mbc{M_{\rm bc}}

\def\DeltaE{\Delta E}
\def\Ebeam{E_{\rm beam}^*}

\def\Mll{M_{\ell\ell}}
\def\Mllgamma{M_{\ell\ell\gamma}}
\def\Mee{M_{ee}}

\def\Cseven{C_7}
\def\Cnine{C_9}
\def\Cten{C_{10}}

\def\Br{{\cal B}}

\def\LRcont{{\cal R}_{\rm cont}}
\def\LRsl{{\cal R}_{\rm sl}}

\def\Lzero{{\cal L}_0}
\def\Lmax{{\cal L}_{\rm max}}

\def\PM#1#2{\,^{+#1}_{-#2}}

\def\etal{\textit{et al.}}
\def\Journal#1#2#3#4{{#1} {\bf #2}, #3 (#4)} % {journal}{vol}{page}{year}
\def\NIMA{Nucl. Instrum. Meth. A}
\def\NPB{Nucl. Phys. B}
\def\PLB{Phys. Lett. B}
\def\PRL{Phys. Rev. Lett.}
\def\PRD{Phys. Rev. D}

\def\EPJD{Eur. Phys. J. direct C}

\begin{document}

%%%%%%%%%%%%%%%%%%%%%%%%%%%%%%%%%%%%%%%%%%%%%%%%%%%%%%%%%%%%%%%%%%%%%%%%

\bellelogo

\preprint{\vbox{
  \preprintA
  \preprintB
  \preprintC
}}

\title{\SomeSpaceIfPreprint
Observation of \boldmath$\BtoKstarll$}

%%% Paper:    B -> K(*)ll
%%% Journal:  Physical Review Letters
%%% Contacts: A. Ishikawa (akimawa@hepl.phys.nagoya-u.ac.jp)
%%%           M. Nakao (mikihiko.nakao@kek.jp)
%%% Non-responding authors or those who said NO are commented out.
%%% ====================================================================
%%% Click the RELOAD button on your web browser to see the updated file.
%%% ====================================================================
%%% Use \input{author} to insert this material into your latex file.
%%%%% Force institutions to appear in alphabetical order when typeset.
%%%\affiliation{Aomori University, Aomori}
\affiliation{Budker Institute of Nuclear Physics, Novosibirsk}
\affiliation{Chiba University, Chiba}
%%%\affiliation{Chuo University, Tokyo}
\affiliation{University of Cincinnati, Cincinnati, Ohio 45221}
\affiliation{University of Frankfurt, Frankfurt}
\affiliation{Gyeongsang National University, Chinju}
\affiliation{University of Hawaii, Honolulu, Hawaii 96822}
\affiliation{High Energy Accelerator Research Organization (KEK), Tsukuba}
\affiliation{Hiroshima Institute of Technology, Hiroshima}
\affiliation{Institute of High Energy Physics, Chinese Academy of Sciences, Beijing}
\affiliation{Institute of High Energy Physics, Vienna}
\affiliation{Institute for Theoretical and Experimental Physics, Moscow}
\affiliation{J. Stefan Institute, Ljubljana}
\affiliation{Kanagawa University, Yokohama}
\affiliation{Korea University, Seoul}
%%%\affiliation{Kyoto University, Kyoto}
\affiliation{Kyungpook National University, Taegu}
\affiliation{Institut de Physique des Hautes \'Energies, Universit\'e de Lausanne, Lausanne}
\affiliation{University of Ljubljana, Ljubljana}
\affiliation{University of Maribor, Maribor}
\affiliation{University of Melbourne, Victoria}
\affiliation{Nagoya University, Nagoya}
\affiliation{Nara Women's University, Nara}
\affiliation{National Kaohsiung Normal University, Kaohsiung}
\affiliation{National Lien-Ho Institute of Technology, Miao Li}
\affiliation{Department of Physics, National Taiwan University, Taipei}
\affiliation{H. Niewodniczanski Institute of Nuclear Physics, Krakow}
\affiliation{Nihon Dental College, Niigata}
\affiliation{Niigata University, Niigata}
\affiliation{Osaka City University, Osaka}
\affiliation{Osaka University, Osaka}
\affiliation{Panjab University, Chandigarh}
\affiliation{Peking University, Beijing}
\affiliation{Princeton University, Princeton, New Jersey 08545}
\affiliation{RIKEN BNL Research Center, Upton, New York 11973}
\affiliation{Saga University, Saga}
\affiliation{University of Science and Technology of China, Hefei}
\affiliation{Seoul National University, Seoul}
\affiliation{Sungkyunkwan University, Suwon}
\affiliation{University of Sydney, Sydney NSW}
\affiliation{Tata Institute of Fundamental Research, Bombay}
\affiliation{Toho University, Funabashi}
\affiliation{Tohoku Gakuin University, Tagajo}
\affiliation{Tohoku University, Sendai}
\affiliation{Department of Physics, University of Tokyo, Tokyo}
\affiliation{Tokyo Institute of Technology, Tokyo}
\affiliation{Tokyo Metropolitan University, Tokyo}
\affiliation{Tokyo University of Agriculture and Technology, Tokyo}
\affiliation{Toyama National College of Maritime Technology, Toyama}
\affiliation{University of Tsukuba, Tsukuba}
\affiliation{Utkal University, Bhubaneswer}
\affiliation{Virginia Polytechnic Institute and State University, Blacksburg, Virginia 24061}
\affiliation{Yokkaichi University, Yokkaichi}
\affiliation{Yonsei University, Seoul}
  \author{A.~Ishikawa}\affiliation{Nagoya University, Nagoya} % Nagoya
  \author{K.~Abe}\affiliation{High Energy Accelerator Research Organization (KEK), Tsukuba} % KEK
  \author{K.~Abe}\affiliation{Tohoku Gakuin University, Tagajo} % TohokuGakuin
% \author{N.~Abe}\affiliation{Tokyo Institute of Technology, Tokyo} % TIT
  \author{T.~Abe}\affiliation{High Energy Accelerator Research Organization (KEK), Tsukuba} % KEK
  \author{I.~Adachi}\affiliation{High Energy Accelerator Research Organization (KEK), Tsukuba} % KEK
  \author{Byoung~Sup~Ahn}\affiliation{Korea University, Seoul} % Korea
  \author{H.~Aihara}\affiliation{Department of Physics, University of Tokyo, Tokyo} % Tokyo
  \author{K.~Akai}\affiliation{High Energy Accelerator Research Organization (KEK), Tsukuba} % KEK
  \author{M.~Akatsu}\affiliation{Nagoya University, Nagoya} % Nagoya
  \author{M.~Akemoto}\affiliation{High Energy Accelerator Research Organization (KEK), Tsukuba} % KEK
% \author{M.~Asai}\affiliation{Hiroshima Institute of Technology, Hiroshima} % Hiroshima
% ...
% Ljubljana
% \author{D.~Z\"urcher}\affiliation{Institut de Physique des Hautes \'Energies, Universit\'e de Lausanne, Lausanne} % Lausanne
\collaboration{The Belle Collaboration}

\mydate

%%%%%%%%%%%%%%%%%%%%%%%%%%%%%%%%%%%%%%%%%%%%%%%%%%%%%%%%%%%%%%%%%%%%%%%%
\begin{abstract} %------------------------------------------------------

We report the first observation of the flavor-changing neutral current
decay $\BtoKstarll$ and an improved measurement of the decay $\BtoKll$,
where $\ell$ represents an electron or a muon, with a data sample of
$140\fbinv$ accumulated at the $\Upsilon(4S)$ resonance with the Belle
detector at KEKB.  The results for the branching fractions are
$\Br(\BtoKstarll)=\ResultBrBtoKstarll$ and
$\Br(\BtoKll)=\ResultBrBtoKll$, where the first error is statistical,
the second is systematic and the third is from model dependence.

\end{abstract}

%%%%%%%%%%%%%%%%%%%%%%%%%%%%%%%%%%%%%%%%%%%%%%%%%%%%%%%%%%%%%%%%%%%%%%%%


% 11.30.Hv = Flavor symmetries
% 13.20.He = Leptonic and semileptonic decays of bottom mesons
% 14.40.Nd = Properties of Bottom mesons
% 14.65.Fy = Properties of Bottom quarks

\maketitle

%%%%%%%%%%%%%%%%%%%%%%%%%%%%%%%%%%%%%%%%%%%%%%%%%%%%%%%%%%%%%%%%%%%%%%%%
% MAIN TEXT %%%%%%%%%%%%%%%%%%%%%%%%%%%%%%%%%%%%%%%%%%%%%%%%%%%%%%%%%%%%

                       %%%%%%%%%%%%%%%%%%%%%%
                       %%%  INTRODUCTION  %%%
                       %%%%%%%%%%%%%%%%%%%%%%

Flavor-changing neutral current (FCNC) processes are forbidden at tree
level in the Standard Model (SM); they only proceed at a low rate via
higher-order loop diagrams.  SM decay amplitudes for the FCNC processes
$\BtoXsgamma$ and $\BtoXsll$, where $\Xs$ denotes inclusive hadronic
final states with a strangeness $S=\pm1$ and $\ell$ represents an
electron or a muon, have been calculated with small
errors~\cite{bib:sm-fcnc}.  If additional diagrams with non-SM particles
contribute to these FCNC processes, their amplitudes will interfere with
the SM amplitudes, making these processes ideal places to search for new
physics~\cite{bib:beyond-sm}.

Measurements of the decay rate for $\BtoXsgamma$~\cite{bib:all-xsgam} as
well as the recent first exclusive and inclusive measurements by Belle
for $\BtoKll$~\cite{bib:belle-kll} and $\BtoXsll$~\cite{bib:belle-xsll}
have so far shown no disagreement with the SM predictions.  Deviations
due to non-SM amplitudes are often expressed in terms of the Wilson
coefficients $\Cseven$, $\Cnine$ and $\Cten$; a strong constraint on the
magnitude of $\Cseven$ has been set by $\BtoXsgamma$, and a large area
of the $\Cnine$--$\Cten$ plane has been excluded by $\BtoKll$ and
$\BtoXsll$~\cite{bib:ph-algh}.  A complete determination of all three
Wilson coefficients, including the sign of $\Cseven$, requires the
measurement of the forward-backward asymmetry in $\BtoKstarll$ or
$\BtoXsll$; however, $\BtoKstarll$ has not been previously observed.

In this Letter, we report the first observation of the decay
$\BtoKstarll$, using a data sample of 152 million $B$ meson pairs,
corresponding to $140\fbinv$ taken at the $\Upsilon(4S)$ resonance.  We
also report an improved measurement of $\BtoKll$, superseding our
previous result based on $29\fbinv$~\cite{bib:belle-kll}.

The data are produced in $\epem$ annihilation at the KEKB
energy-asymmetric (3.5 on 8 GeV) collider~\cite{bib:kekb} and collected
with the Belle detector~\cite{bib:belle-detector}.  The Belle detector
is a large-solid-angle spectrometer that includes a three-layer silicon
vertex detector (SVD), a 50-layer central drift chamber (CDC), an array
of aerogel threshold Cherenkov counters (ACC), time-of-flight (TOF)
scintillation counters, and an electromagnetic calorimeter (ECL)
comprised of CsI(Tl) crystals located inside a superconducting solenoid
coil that provides a 1.5 T magnetic field.  An iron flux-return located
outside of the coil is instrumented to identify muons (KLM).

                      %%%%%%%%%%%%%%%%%%%%%%%%
                      %%%  RECONSTRUCTION  %%%
                      %%%%%%%%%%%%%%%%%%%%%%%%

The event reconstruction procedure is similar to our previous
report~\cite{bib:belle-kll}.  We reconstruct the following final states:
$\BtoKstarZll$, $\BtoKstarPll$, $\BtoKSll$ and $\BtoKPll$.  Charge
conjugate modes are implied throughout this Letter.  The following decay
chains are used to reconstruct the intermediate states:
$\KstZ\to\KP\piM$, $\KstP\to\KS\piP$ and $\KstP\to\KP\piZ$,
$\KS\to\piP\piM$, and $\piZ\to\gamma\gamma$.

Charged tracks are classified as $e$, $\mu$, $K$ and $\pi$ candidates by
discriminating between the flavors for the pairwise combinations, using
criteria which allow multiple classifications of an individual track.
The $e/h$ discriminant (where $h = K$ or $\pi$) is formed from the
energy deposit in the ECL, the specific ionization measurements in the
CDC, and the ACC light yield.  The $\mu/h$ discriminant is based on the
hits in the KLM.  The $K/\pi$ and $K/\mu$ discriminants use the CDC,
ACC, and TOF information.  Electrons, muons and kaons are selected using
loose conditions on the $e/h$, $\mu/h$ and $K/\pi$ discriminants,
respectively.  All tracks are classified as pions unless they satisfy
tight conditions on $e/h$ or $K/\pi$; the same $e/h$ condition is
required for kaons, and a similar $K/\mu$ condition is required for
muons.  To reduce the misidentification of hadrons as leptons, we
require minimum momenta of $0.4\GeVc$ and $0.7\GeVc$ for electrons and
muons, respectively.  We apply a tight requirement for the muons below
$1.0\GeVc$.  Each of the charged tracks, except for the $\KS\to\piP\piM$
daughters, is required to have an impact parameter with respect to the
interaction point of less than 0.5~cm transverse to, and 5.0~cm along
the positron beam axis.  Photons are reconstructed within the ECL with a
minimum energy requirement of $50\MeV$.

Invariant masses for the $\piZ$, $\KS$ and $K^*$ candidates are required
to be within windows of $\pm10\MeVcc$ (${\sim}2\sigma$), $\pm15\MeVcc$
(${\sim}3.3\sigma$) and ${\pm}75\MeVcc$ ($1.5\Gamma$), respectively,
around their nominal masses.  We require a minimum momentum of
$0.1\GeVc$ for the $\piZ$ candidates.  We impose $\KS$ selection
criteria based on the distance and the direction of the $\KS$ vertex and
the impact parameters of daughter tracks.  For $K^{*+}\to\KP\piZ$,
$\cos\theta_{\rm{hel}}<0.8$ is required to reduce background from soft
$\piZ$s, where $\theta_{\rm{hel}}$ is the angle between the $K^{*+}$
momentum in the $B$ rest frame and the $K^+$ momentum in the $K^{*+}$
rest frame.

We form $B$ candidates by combining a $K^{(*)}$ candidate and an
oppositely charged lepton pair using two variables: the beam-energy
constrained mass $\Mbc = \sqrt{ (\Ebeam/c^2)^2 - |p_{B}^*/c|^{2}}$ and
the energy difference $\Delta E = E^*_{B} - \Ebeam$, where $p^*_{B}$ and
$E^*_B$ are the measured momentum and energy, respectively, of the $B$
candidate, and $\Ebeam$ is the beam energy.  Here and throughout this
Letter, variables denoted with an asterisk are calculated in the
$\Upsilon(4S)$ rest frame.  When multiple candidates are found in an
event, we select the candidate with the smallest value of $|\DeltaE|$.

                        %%%%%%%%%%%%%%%%%%%%
                        %%%  BACKGROUND  %%%
                        %%%%%%%%%%%%%%%%%%%%

The following five types of backgrounds are considered.  1)~{\it
Charmonium} $B$ decay background from $B\to J/\psi X_s$ and
$B\to\psi'X_s$ decays is removed by vetoing lepton pairs whose invariant
mass ($\Mll$) is near the $J/\psi$ or $\psi'$ mass~\cite{bib:belle-kll}.
In addition, we reject events that have a photon with energy less than
$500\MeV$ within a $50\mrad$ cone around either the electron or positron
direction (or a photon within each cone) and an $\epem\gamma(\gamma)$
invariant mass within the veto windows.  For $\Kst\elel$ modes, we
reject the event if an unobserved photon along one of the lepton
directions with an energy $\Ebeam-E_K^*-E^*_{\ell\ell}$ can replace the
pion, giving $\Mllgamma$ and $\Mbc$ consistent with $\BtoJpsiK$.  2)~We
suppress background from {\it photon conversions} and
$\piZ\to\epem\gamma$ by requiring the dielectron mass to satisfy $\Mee >
0.14\GeVcc$. This eliminates possible background from $\BtoKstargamma$
and $\BtoKorKstarPZ$.  3)~Background from {\it continuum} $\qqbar$
($q=u,d,s,c$) production is suppressed using a likelihood ratio
$\LRcont$ formed from a Fisher discriminant, $\cos\theta^*_B$, and, for
$K^{(*)}\epem$ only, $\cos\theta^*_{\rm sph}$.  The Fisher
discriminant~\cite{bib:fisher} is calculated from the energy flow in 9
cones along the $B$ candidate sphericity axis and the normalized second
Fox-Wolfram moment $R_{2}$~\cite{bib:fox-wolfram}.  The angles
$\theta^*_B$ and $\theta^*_{\rm sph}$ are the $B$ meson angles with
respect to the beam and the sphericity axes, respectively.  4)~{\it
Semileptonic} $B$ decay background is suppressed using another
likelihood ratio $\LRsl$, formed from the missing energy of the event,
$E^*_{\mathrm{miss}}$, and $\cos\theta^*_B$.  5)~{\it Hadronic} $B$
decay background, $B\to K^{(*)}h^+h^-$, e.g., from $B\to D\pi$, can
contribute if two hadrons are misidentified as leptons.  We find that
other potential backgrounds are negligible.

                       %%%%%%%%%%%%%%%%%%%%%%
                       %%%  OPTIMIZATION  %%%
                       %%%%%%%%%%%%%%%%%%%%%%

For each decay mode, the selection criteria on the two likelihood ratios
$\LRcont$ and $\LRsl$ are chosen to maximize $N_S/\sqrt{N_S+N_B}$, where
$N_S$ is the expected signal yield and $N_B$ is the expected background
in the $\Mbc$ and $\DeltaE$ signal windows.  The signal windows
(${\sim}2.5\sigma$) are defined as $|\Mbc-M_B| <0.007\GeVcc$ for both
lepton modes and $-0.055 (-0.035) \GeV<\DeltaE<0.035\GeV$
% ($|\DeltaE|<0.035\GeV$) 
for the electron (muon) mode.  A large Monte Carlo (MC) background
sample of a mixture of $b\to c$ decays and $\epem\to\qqbar$ events is
used to estimate $N_B$.  The $\BtoKorKstarll$ signal events are
generated according to Ref.~\cite{bib:ph-algh} to determine $N_S$, and
to estimate the efficiencies that are summarized in
Table~\ref{tab:results}.

                         %%%%%%%%%%%%%%%%%
                         %%%  FITTING  %%%
                         %%%%%%%%%%%%%%%%%

The signal yield is determined by a binned maximum-likelihood fit to the
$\Mbc$ distribution for the events within the $\DeltaE$ signal window
using a Gaussian signal plus three background functions.  The area of
this Gaussian function is the signal yield; the mean and width are
determined using observed $\JpsiKorKstar$ events.  We find no dilepton
mass dependence of the width and mean using a MC study.  The first
background function is for the semileptonic $B$ decays and, to a lesser
extent, the continuum background, and is modeled with a threshold
function~\cite{bib:argus-function} whose shape parameter is determined
using a large MC sample that contains oppositely charged leptons and
whose normalization is allowed to float.  This MC sample reproduces the
background parametrization for $\BtoKorKstaremu$ data in which only
combinatorial background is expected.  
The two other background functions account for the residual $B$ to
charmonium decays and hadronic $B$ decays, and are modeled with separate
combinations of a similar threshold function and an additional Gaussian
component.
The shape and the size of the charmonium background function are fixed
from $J/\psi$ and $\psi'$ inclusive MC samples.  We find the Gaussian
component of the charmonium background contributes less than one event.
The shape and the size of the hadronic background are evaluated using
hadron enriched data by relaxing the lepton identification criteria.
The Gaussian components of the hadronic background contribution,
multiplied by the lepton misidentification probability (measured in
bins of momentum and polar angle with respect to the positron beam), are
then found to be $1.05\pm0.08$ and $0.64\pm0.05$ events for $\BtoKll$ and
$\BtoKstarll$, respectively.

                          %%%%%%%%%%%%%%%
                          %%%  YIELD  %%%
                          %%%%%%%%%%%%%%%

\begin{figure}[ht]
  \includegraphics[scale=\figonescale]{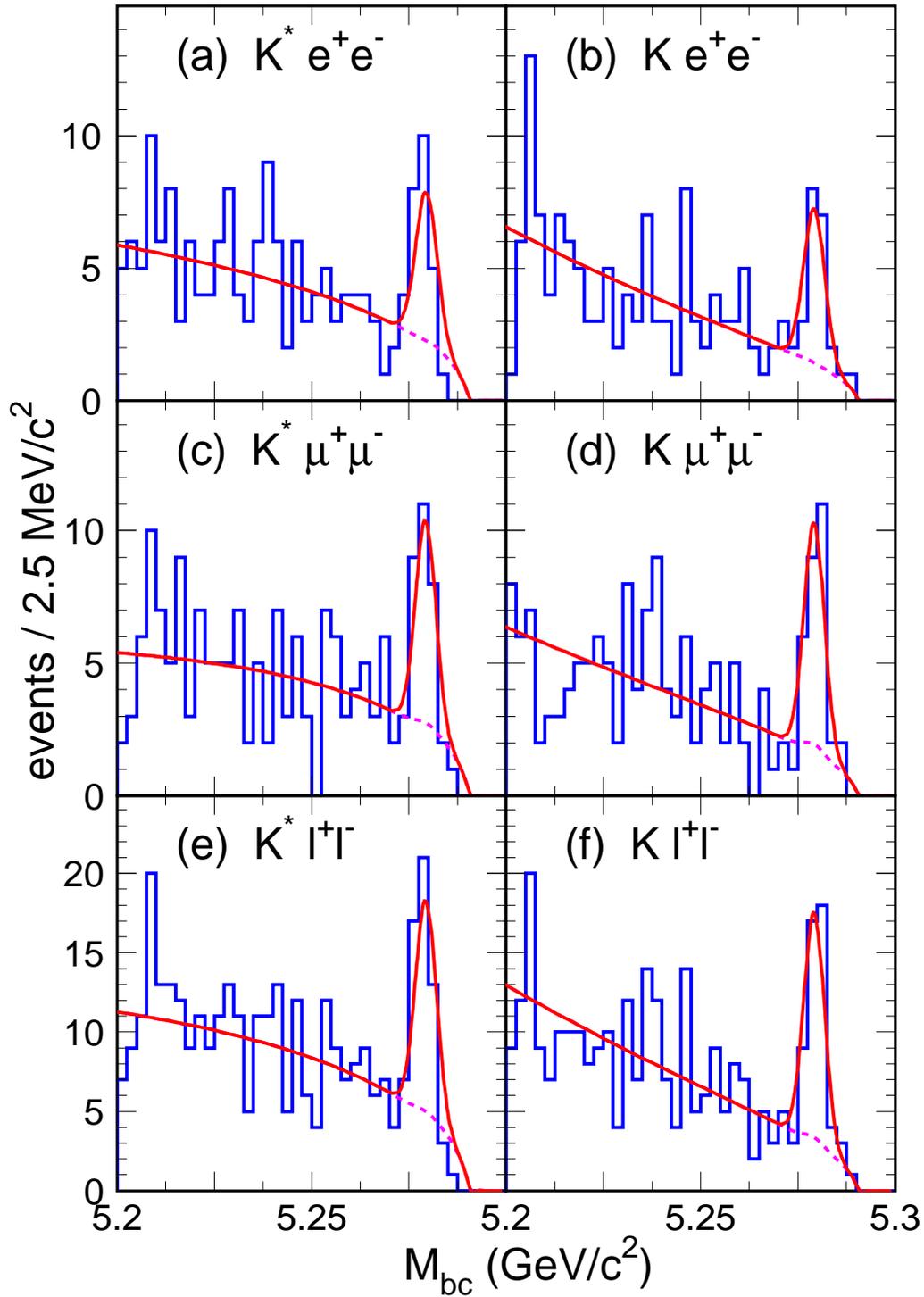}
  \caption{%
    $M_{\rm bc}$ distributions (histograms) for $K^{(*)}\elel$
    samples.  Solid and dotted curves show the results of the fits
    and the background contributions, respectively.}
  \label{fig:mbcfit}
\end{figure}

Figure~\ref{fig:mbcfit} and Table~\ref{tab:results} give the fit
results.  We observe
$35.8^{+8.0}_{-7.3}{\rm(stat.)}{\pm}1.7{\rm(syst.)}$ $\BtoKstarll$
signal events with a significance of $5.7$, and
$37.9^{+7.6}_{-6.9}{\rm(stat.)}^{+1.0}_{-1.1}{\rm(syst.)}$ $\BtoKll$
signal events with a significance of $7.4$.  The error due to
uncertainty in the fixed parameters is included in the systematic error.
To evaluate the uncertainty in the signal function parametrization, the
mean and width of the Gaussian function are changed by $\pm 1$ standard
deviation ($\sigma$) from the values determined from $J/\psi K^{(*)}$
events.  The uncertainty in the semileptonic plus continuum background
parametrization, which is the largest error source, is obtained by
varying the parameter by $\pm 1 \sigma$ from the value determined with a
large MC sample. The uncertainties of the hadronic (charmonium)
background contributions are evaluated by changing the shape parameters
and the normalizations of the Gaussian and threshold components by
$\pm1\sigma$ ($\pm100\%$).  The significance is defined as
$\sqrt{-2\ln(\Lzero/\Lmax)}$, where $\Lmax$ is the maximum likelihood in
the $\Mbc$ fit and $\Lzero$ is the likelihood of the best fit when the
signal yield is constrained to be zero.  In order to include the effect
of systematic error in the significance calculation, we use the
parameters simultaneously changed by $1\sigma$ ($100\%$ for the
charmonium background) in the direction that reduces the resulting
significance.

                       %%%%%%%%%%%%%%%%%%%%%
                       %%%  SYSTEMATICS  %%%
                       %%%%%%%%%%%%%%%%%%%%%

In addition to the systematic error in the signal yield, we consider the
following experimental systematic errors in the efficiency
determination.  For each charged track, we estimate the systematic error
due to reconstruction efficiency to be 1.0\%, and the systematic errors
due to kaon, pion, electron and muon identification to be 1.0\%, 0.8\%,
0.5\% and 1.2\%, respectively.  For each $\KS$ candidate and $\piZ$
candidate, we estimate the systematic errors due to reconstruction
efficiencies to be 4.5\% and 2.7\%, respectively.  The uncertainty in
the background suppression is estimated to be 2.3\% using $J/\psi
K^{(*)}$ control samples.  Systematic errors due to MC statistics range
from 0.5\% to 2.2\%.  All these errors are added in quadrature.

                       %%%%%%%%%%%%%%%%%%%%%
                       %%%  MODEL ERROR  %%%
                       %%%%%%%%%%%%%%%%%%%%%

The uncertainty due to the theoretical model assumptions is evaluated by
calculating the efficiency for signal MC samples generated using three
form-factor models~\cite{bib:ph-algh,bib:kll-twomore} and taking the
maximum difference as the model-dependence error.

                    %%%%%%%%%%%%%%%%%%%%%%%%%%%%
                    %%%  BRANCHING FRACTION  %%%
                    %%%%%%%%%%%%%%%%%%%%%%%%%%%%

When calculating the branching fractions, we assume an equal production
rate for charged and neutral $B$ meson pairs, isospin invariance, lepton
universality for $\BtoKll$, and the branching ratio
$\Br(\BtoKstaree)/\Br(\BtoKstarmumu)=1.33$~\cite{bib:ph-algh}. The
combined efficiency and branching fraction are scaled to the muon mode.
We find
\[
  \begin{array}{r@{\;=\;}l}
  {\displaystyle\Br(\BtoKstarll)} & {\displaystyle\ResultBrBtoKstarll}, \\
  {\displaystyle\Br(\BtoKll)}     & {\displaystyle\ResultBrBtoKll}, \\
  \end{array}
\]
where the first error is statistical, the second is systematic, and the
third is from model dependence.  This systematic error is a quadratic
sum of the systematic errors in the yield and efficiency, and the
uncertainty in $B$ meson pair counting of 0.5 \%.  The results are
consistent with the SM
predictions~\cite{bib:ph-algh,bib:kll-twomore,bib:kll-others}, our
previous values~\cite{bib:belle-kll}, and results recently reported by
BaBar~\cite{bib:babar-kll}.  The complete set of results is given in
Table~\ref{tab:results}.

                       %%%%%%%%%%%%%%%%%%%%%%
                       %%%  UPPER LIMITS  %%%
                       %%%%%%%%%%%%%%%%%%%%%%

For the modes with a significance of less than 3, we set 90\% confidence
level upper limits.  The upper limit on the yield, $N$, is defined as
$\int^N_0 {\cal{L}}(n)dn = 0.9 \int^{\infty}_0 {\cal{L}}(n)dn$.  The
function ${\cal{L}}(n)$ is the likelihood for signal yield $n$, using
signal and background shape parameters that are modified by $1\sigma$ of
their errors in the direction to increase the signal yield.  The upper
limits for the branching fractions are then calculated by using the
efficiencies reduced by $1\sigma$ of their errors.

                     %%%%%%%%%%%%%%%%%%%%%%%%%
                     %%%  Q2 DISTRIBUTION  %%%
                     %%%%%%%%%%%%%%%%%%%%%%%%%

\begin{figure}[ht]
  \includegraphics[scale=\figtwoscale]{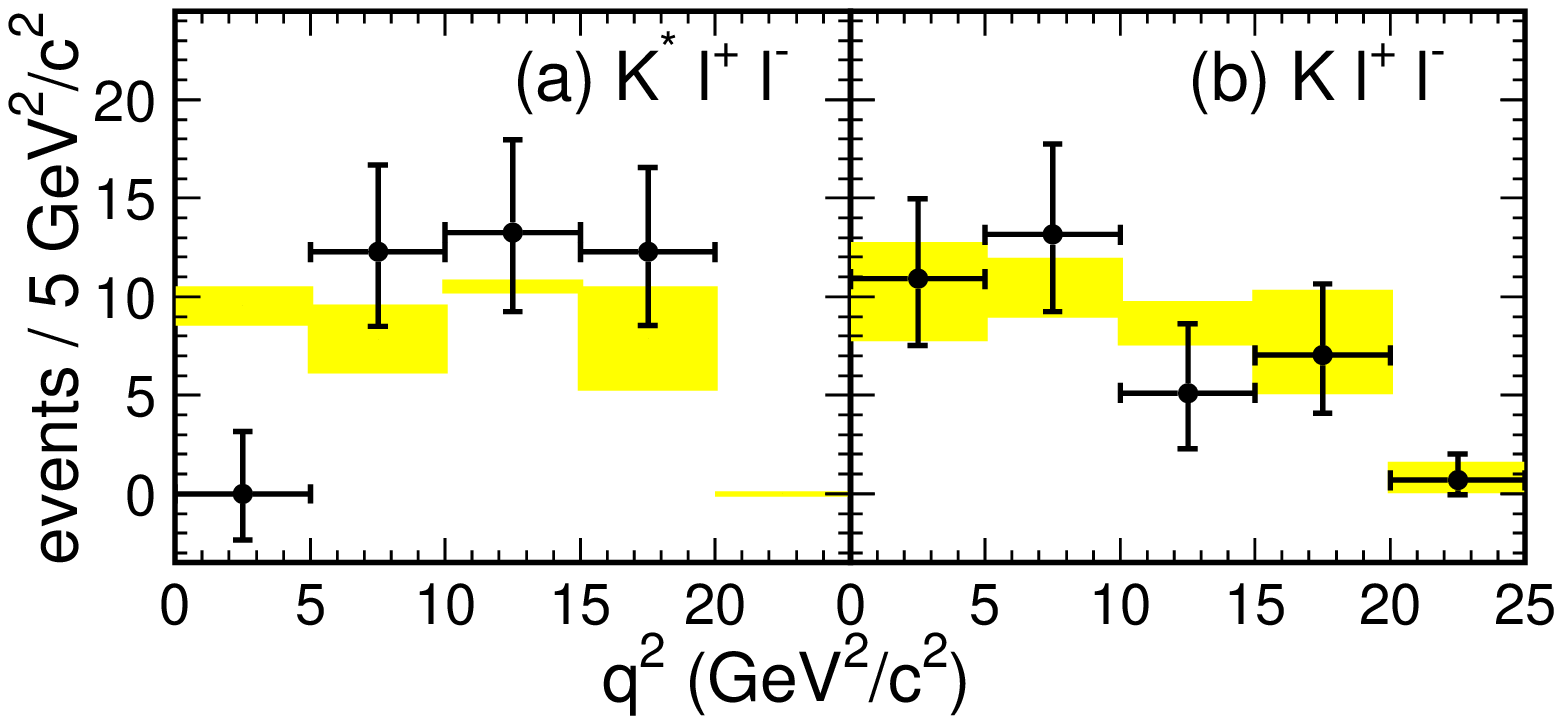}
  \caption{%
    $q^2$ distributions of (a) $\Kstarll$ and (b) $\Kll$. Points
    with error bars show the data while the hatched boxes show the range
    of SM expectations from various
    models~\cite{bib:ph-algh,bib:kll-twomore}.  Statistical and
    systematic errors are added in quadrature.}
  \label{fig:q2}
\end{figure}

Figure~\ref{fig:q2} shows the measured $q^2=\Mll^2c^2$ distributions
for $\BtoKll$ and $\Kstarll$.  The signal yield is extracted in each
$q^2$ bin from a fit to the $\Mbc$ distributions.

                          %%%%%%%%%%%%%%%%%
                          %%%  SUMMARY  %%%
                          %%%%%%%%%%%%%%%%%

In summary, we have observed the decay $\BtoKstarll$ for the first time.
This mode will provide a useful sample for a forward-backward asymmetry
measurement.  The $\BtoKll$ decay is also measured with improved
accuracy.  The measured branching fractions are in agreement with the SM
predictions, and may be used to provide more stringent constraints on
physics beyond the SM.

%%%%%%%%%%%%%%%%%%%%%%%%%%%%%%%%%%%%%%%%%%%%%%%%%%%%%%%%%%%%%%%%%%%%%%%%
%%%%%%%%%%%%%%%%%%%%%%%%%%%%%%%%%%%%%%%%%%%%%%%%%%%%%%%%%%%%%%%%%%%%%%%%
%%%%%%%%%%%%%%%%%%%%%%%%%%%%%%%%%%%%%%%%%%%%%%%%%%%%%%%%%%%%%%%%%%%%%%%%

\begin{table*}[ht]
  \caption{%
    Summary of the results: signal yields obtained from the
    $M_{\mathrm{bc}}$ fit and their significances, reconstruction
    efficiencies including the intermediate branching fractions,
    branching fractions ($\Br$) and their 90\% confidence level upper
    limits. }
  \label{tab:results}
  \begin{ruledtabular}
  \begin{tabular}{lcccccc}
  %%%%%%%%%%%%%%%%%%%%%%%%%%%%%%%%%%%%%%%%%%%%%%%%%%%%%%%%%%%%%%%%%%%%%%%%%%%
  Mode         & Signal yield              & Significance & Efficiency [\%]
               & ${\cal{B}}$ [$\times10^{-7}$]    &  Upper Limit [$\times10^{-7}$] \\
  {}           & $\pm$stat.$\pm$syst.      &      &  $\pm$syst.$\pm$model
               & $\pm$stat.$\pm$syst.$\pm$model   &                        \\
  \hline %%%%%%%%%%%%%%%%%%%%%%%%%%%%%%%%%%%%%%%%%%%%%%%%%%%%%%%%%%%%%%%%%%%%
  $\KstZ\epem$ & $10.2\PM{4.5}{3.8}\pm0.8$        &  2.8  & $5.2\pm0.3\pm0.04$ 
               & $12.9\PM{5.7}{4.9}\pm{1.1}\pm0.1$        & 24 \\
  $\KstP\epem$ &  $5.3\PM{3.3}{2.6}\PM{0.5}{0.6}$ &  1.9  & $1.7\pm0.1\pm0.1$
               & $20.2\PM{12.7}{10.1}\PM{2.3}{2.4}\pm0.7$ & 46 \\
  $\Kst\epem$  & $15.6\PM{5.5}{4.8}\pm1.0$        &  3.5  & $3.5\pm0.2\pm0.04$
               & $14.9\PM{5.2}{4.6}\PM{1.2}{1.3}\pm0.2$   & --- \\
  \hline %%%%%%%%%%%%%%%%%%%%%%%%%%%%%%%%%%%%%%%%%%%%%%%%%%%%%%%%%%%%%%%%%%%%
  $\KZ\epem$   & $0.0\PM{1.5}{0.9}\PM{0.2}{0.3}$  &  0.0  & $5.0\pm0.3\pm0.1$
               & $0.0\PM{2.0}{1.2}\PM{0.3}{0.4}\pm0.0$    &  5.4 \\
  $\KP\epem$   & $15.9\PM{4.9}{4.2}\pm0.6$        &  5.1  & $16.6\pm0.7\pm0.4$
               &  $6.3\PM{1.9}{1.7}\pm0.3\pm0.1$          &  --- \\
  $K\epem$     & $15.9\PM{5.1}{4.4}\pm0.7$        &  4.5  & $10.8\pm0.5\pm0.2$
               &  $4.8\PM{1.5}{1.3}\pm0.3\pm0.1$          &  --- \\
  \hline %%%%%%%%%%%%%%%%%%%%%%%%%%%%%%%%%%%%%%%%%%%%%%%%%%%%%%%%%%%%%%%%%%%%
  \hline %%%%%%%%%%%%%%%%%%%%%%%%%%%%%%%%%%%%%%%%%%%%%%%%%%%%%%%%%%%%%%%%%%%%
  $\KstZ\mumu$ & $17.1\PM{5.4}{4.7}\pm0.9$        &  4.2  & $8.5\pm0.5\pm0.3$
               & $13.3\PM{4.2}{3.7}\pm1.0\pm0.5$          &  --- \\
  $\KstP\mumu$ &  $2.8\PM{2.9}{2.3}\pm0.6$        &  0.8  & $2.8\pm0.2\pm0.2$
               &  $6.5\PM{6.9}{5.3}\PM{1.4}{1.5}\pm0.4$   &  22  \\ 
  $\Kst\mumu$  & $20.0\PM{6.0}{5.3}\PM{1.1}{1.2}$ &  4.2  & $5.6\pm0.3\pm0.2$
               & $11.7\PM{3.6}{3.1}\pm0.9\pm0.5$          &  --- \\
  \hline %%%%%%%%%%%%%%%%%%%%%%%%%%%%%%%%%%%%%%%%%%%%%%%%%%%%%%%%%%%%%%%%%%%%
  $\KZ\mumu$   &  $5.7\PM{3.0}{2.3}\PM{0.2}{0.3}$ &  3.1  & $6.7\pm0.4\pm0.3$
               &  $5.6\PM{2.9}{2.3}\pm0.4\pm0.3$          &  --- \\
  $\KP\mumu$   & $16.3\PM{5.1}{4.5}\PM{0.7}{0.8}$ &  4.6  & $23.6\pm1.1\pm0.6$
               &  $4.5\PM{1.4}{1.2}\pm0.3\pm0.1$          &  --- \\ 
  $K\mumu$     & $22.0\PM{5.8}{5.1}\pm0.8$        &  5.6  & $15.2\pm0.7\pm0.5$
               &  $4.8\PM{1.2}{1.1}\pm0.3\pm0.2$          &  --- \\
  \hline %%%%%%%%%%%%%%%%%%%%%%%%%%%%%%%%%%%%%%%%%%%%%%%%%%%%%%%%%%%%%%%%%%%%
  \hline %%%%%%%%%%%%%%%%%%%%%%%%%%%%%%%%%%%%%%%%%%%%%%%%%%%%%%%%%%%%%%%%%%%%
  $\KstZ\elel$ & $27.4\PM{6.9}{6.2}\pm1.3$        &  5.2  &  $7.7\pm0.4\pm0.2$
               & $11.7\PM{3.0}{2.7}\pm0.8\pm0.3$          &  ---  \\
  $\KstP\elel$ &  $8.1\PM{4.3}{3.3}\PM{0.8}{0.9}$ &  2.1  &  $2.5\pm0.2\pm0.05$
               & $10.5\PM{5.6}{4.3}\PM{1.2}{1.1}\pm0.2$   &  22  \\ 
  $\Kst\elel$  & $35.8\PM{8.0}{7.3}\pm1.7$        &  5.7  &  $5.1\pm0.3\pm0.1$
               & $11.5\PM{2.6}{2.4}\pm0.8\pm0.2$          &  ---  \\
 \hline %%%%%%%%%%%%%%%%%%%%%%%%%%%%%%%%%%%%%%%%%%%%%%%%%%%%%%%%%%%%%%%%%
  $\KZ\elel$   &  $5.7\PM{3.4}{2.7}\PM{0.4}{0.5}$ &  2.3  & $5.9\pm0.4\pm0.2$
               &  $3.2\PM{1.9}{1.5}\pm0.3\pm0.1$          &  6.8 \\ 
  $\KP\elel$   & $32.3\PM{6.9}{6.2}\PM{0.9}{1.0}$ &  7.0  & $20.1\pm0.9\pm0.1$
               &  $5.3\PM{1.1}{1.0}\pm0.3\pm0.04$         &  --- \\ 
  $K\elel$     & $37.9\PM{7.6}{6.9}\PM{1.0}{1.1}$ &  7.4  & $13.0\pm0.6\pm0.2$
               &  $4.8\PM{1.0}{0.9}\pm0.3\pm0.1$          &  --- \\
  %%%%%%%%%%%%%%%%%%%%%%%%%%%%%%%%%%%%%%%%%%%%%%%%%%%%%%%%%%%%%%%%%%%%%%%%%%%
  \end{tabular}
  \end{ruledtabular}
\end{table*}

%%%%%%%%%%%%%%%%%%%%%%%%%%%%%%%%%%%%%%%%%%%%%%%%%%%%%%%%%%%%%%%%%%%%%%%%
% ACKNOWLEDGEMENTS -----------------------------------------------------

We wish to thank the KEKB accelerator group for the excellent
operation of the KEKB accelerator.
We acknowledge support from the Ministry of Education,
Culture, Sports, Science, and Technology of Japan
and the Japan Society for the Promotion of Science;
the Australian Research Council
and the Australian Department of Education, Science and Training;
the National Science Foundation of China under contract No.~10175071;
the Department of Science and Technology of India;
the BK21 program of the Ministry of Education of Korea
and the CHEP SRC program of the Korea Science and Engineering Foundation;
the Polish State Committee for Scientific Research
under contract No.~2P03B 01324;
the Ministry of Science and Technology of the Russian Federation;
the Ministry of Education, Science and Sport of the Republic of Slovenia;
the National Science Council and the Ministry of Education of Taiwan;
and the {U.S.} Department of Energy.

%%%%%%%%%%%%%%%%%%%%%%%%%%%%%%%%%%%%%%%%%%%%%%%%%%%%%%%%%%%%%%%%%%%%%%%%
% REFERENCES -----------------------------------------------------------

%%%%%%%%%%%%%%%%%%%%%%%%%%%%%%%%%%%%%%%%%%%%%%%%%%%%%%%%%%%%%%%%%%%%%%%%
% FIGURES AND TABLES ---------------------------------------------------

\end{document}